\documentclass[twocolumn, pra, showpacs,superscriptaddress]{revtex4}
\usepackage{amsfonts}
\usepackage{amsmath}
\usepackage{amssymb}
\usepackage{graphicx}

\begin{document}

\title{Magnetization Oscillation of a Spinor Condensate Induced by
Magnetic Field Gradient}
\author{Jie Zhang}
\affiliation{Institute of Theoretical Physics, Shanxi University, Taiyuan 030006,
People's Republic of China}
\author{Baoguo Yang}
\affiliation{Institute of Theoretical Physics, Shanxi University, Taiyuan 030006,
People's Republic of China}
\author{Yunbo Zhang}
\email{ybzhang@sxu.edu.cn}
\affiliation{Institute of Theoretical Physics, Shanxi University, Taiyuan 030006,
People's Republic of China}

\begin{abstract}
We study the spin mixing dynamics of ultracold spin-1 atoms in a weak
non-uniform magnetic field with field gradient $G$, which can flip the spin
from $+1$ to $-1$ so that the 
magnetization $m=\rho _{+}-\rho _{-}$ is not any more a constant.
The dynamics of $m_F=0$ Zeeman component $\rho _{0}$, as well as the
system magnetization $m$, are illustrated for both
ferromagnetic and polar interaction cases in the mean-field theory. We find
that the dynamics of system magnetization can be tuned between the
Josephson-like oscillation similar to the case of double well, and the
interesting self-trapping regimes, i.e. the spin mixing dynamics sustains a
spontaneous magnetization. Meanwhile the dynamics of $\rho _0$ may be
sufficiently suppressed for initially imbalanced number distribution in the
case of polar interaction. A "beat-frequency" oscillation of the magnetization 
emerges in the case of balanced initial distribution for polar interaction,
which vanishes for ferromagnetic interaction.
\end{abstract}

\maketitle

\section{Introduction}

Since the successful realization of $^{23}$Na condensate in the optical trap \cite%
{Stamper-Kurn}, with the spin degrees of freedom liberated, the coherent
spin-mixing dynamics inside a spin-1 BEC has been studied intensively \cite%
{Law,Pu,MSChang,MSChang2,Cheng,Lee,WXZhang10}. The spin mixing interaction
allows for exchanging atoms among spin components but conserving the total
angular momentum. Two atoms in Zeeman state $\left\vert
0\right\rangle $ can coherently scatter into the states $\left\vert
1\right\rangle $ and $\left\vert -1\right\rangle ,$ and vice versa: $%
2\left\vert 0\right\rangle \rightleftharpoons \left\vert 1\right\rangle
+\left\vert -1\right\rangle $. As one of the most active topics in quantum gases, such
spin-exchang dynamics was first studied by Law \textit{et. al.}
\cite{Law} and has been observed in the way of population oscillations of the
Zeeman states inside $^{87}$Rb condensates \cite{MSChang2}, where
atoms interact ferromagnetically. A temporal modulation of spin exchange
interaction, which is tunable with optical Feshbach resonance, was recently
proposed to localize the spin mixing dynamics in $^{87}$Rb condensate \cite%
{WXZhang10}.

The properties of a three-component ($F=1$) spinor condensate are first
studied by Ho \cite{Ho} and Ohmi \cite{Ohmi}. For a spin-1 system, atom-atom
interaction takes the form $V(\mathbf{r})=\delta (\mathbf{r})(c_{0}+c_{2}%
\mathbf{F}_{1}\cdot \mathbf{F}_{2})$, where $\mathbf{r}$ is the distance
vector between two atoms, and $c_{0},c_{2}$ denote spin-independent and
spin-exchange interaction respectively. Many predictions are
verified  experimentally \cite{Stenger,domain}, and the most fundamental property concerns
the existence of two different phases determined by $c_{2}$: the so-called
polar ($c_{2}>0$) and ferromagnetic ($c_{2}<0$) states, 
corresponding to the $F=1$ state of $^{23}$Na and $^{87}$Rb atomic
condensates respectively. The fragmented condensate in a uniform magnetic field can be
turned into a single condensate state by a field gradient \cite%
{HoYip,Mueller06}. More exotic ground state phases both in the mean field
level and the fully quantum many body theory have been extended to
condensates with higher spins \cite{Ciobanu,Ueda,Ho3} and recently to spinor
mixtures \cite{Xuone,Xutwo,zj}.


In this paper we study the dependence of the spin dynamics on a small magnetic field
gradient, which practically provides a process to flip the spin between $+1$ 
and $-1$ states thus
turns a fragmented condensate into a coherent one. We adopt the mean-field
approximation, in which a spinor condensate is described by a multi-component
vector field. It has provided surprisingly good descriptions for most
properties of the spinor condensate as evidenced by the experimental
verification of many predictions \cite{MSChang}. As the spin flipping term
is considered in the condensate, the system magnetization exhibits obvious
macroscopic oscillation similar to the Josephson oscillation of a scalar
condensate in a double well \cite{doublewell} and we find that the dynamics of
spin-$0$ component $\rho _{0}$ may be greatly suppressed in the case of
polar interaction. This provides us an intriguing tool to manipulate the
atomic population in spinor condensate.

\section{The effective Hamiltonian for the system}

The many-body Hamiltonian of $N$ spin-1 atoms of mass $M$ in a uniform
magnetic field reads%
\begin{equation}
H=\sum\limits_{k=1}^{N}(\frac{\mathbf{P}_{k}^{2}}{2M}+V_{trap}+\gamma
\mathbf{B}_{0}\cdot \mathbf{F}_{k})+\sum_{k<l}V_{int}(\mathbf{r}_{k}-\mathbf{%
r}_{l}).  \label{hamzero}
\end{equation}%
The atoms are loaded into an optical trap $V_{trap}$ and $V_{int}$ denotes
the collisional interaction between atoms. $\mathbf{P}_{k}$ and $\mathbf{F}%
_{k}$ are the momentum operators and spin-1 operators of the $k$-th atom,
respectively. We consider a field gradient $G$ as has already been applied
in the MIT experiment \cite{Stenger}, and to be more specific we just replace $%
\mathbf{B}_{0}$ in (\ref{hamzero}) with
\begin{equation}
\mathbf{B}(\mathbf{r})=B_{0}\mathbf{r_{B}}=B_{0}[\mathbf{\hat{z}}+G(x\mathbf{%
\hat{x}}-z\mathbf{\hat{z}})].
\end{equation}%
We choose the local field direction $\mathbf{%
r_{B}}$ as the spin quantization axis \cite{HoYip} by performing a unitary 
transformation $U=\prod\limits_{k=1}^{N}e^{-\frac{i}{%
\hbar }\mathbf{n}(\mathbf{r}_{k})\cdot \mathbf{F}_{k}}$ on the
Hamiltonian (\ref{hamzero})  where $\mathbf{n }=%
\mathbf{\hat{z}\times \mathbf{r_{B}}=}Gx\mathbf{\hat{y}}$. The contact
interaction $V_{int}$ is spin conserving, hence it is
invariant under the transformation. On the other hand, an additional Berry
phase associated with the local change of basis emerges in the momentum
term. Take the single particle hamiltonian as an example, $h=\frac{\mathbf{P}%
^{2}}{2M}+V_{trap}+\gamma \mathbf{B}\cdot \mathbf{F}$, and $U=e^{-\frac{i}{%
\hbar }GxF_{y}}$, we find that
\begin{eqnarray}
U^{\dag }\mathbf{P}^{2}U &=&(P_{x}-A)^{2}+P_{y}^{2}+P_{z}^{2}  \notag \\
&=&\mathbf{P}^{2}-2GF_{y}P_{x}+G^{2}F_{y}^{2}  
\label{p2}
\end{eqnarray}%
where the operator $A=GF_{y}$. The Zeeman energy term is
transformed into \cite{HoShenoy}
\begin{eqnarray}
U^{\dag }\mathbf{B}\cdot \mathbf{F}U &=&B_{0}F_{x}\left[ Gx\cos \left(
Gx\right) -\left( 1-Gz\right) \sin \left( Gx\right) \right]   \notag \\
&+&B_{0}F_{z}\left[ Gx\sin \left( Gx\right) +\left( 1-Gz\right) \cos \left(
Gx\right) \right] . \notag \\
\end{eqnarray}%
For small field gradient $G$, we approximate the trigonometric functions up to
order of $O(G^{2})$ in accordance with Eq. (\ref{p2}), $\cos \left(
Gx\right) \simeq 1+G^{2}x^{2}/2,\sin \left( Gx\right) \simeq Gx$, which
leads us to
\begin{equation}
U^{\dag }\mathbf{B}\cdot \mathbf{F}U=B_{0}[F_{z}+CG+DG^{2}+O(G^{3})]
\end{equation}%
where the operator $%
C=-zF_{z},D=xzF_{x}+x^{2}F_{z}/2$.

The many-body Hamiltonian is finally transformed into
\begin{eqnarray}
H_{eff} &=&\sum\limits_{k=1}^{N}\{\frac{\mathbf{P}_{k}^{2}}{2M}-\frac{G%
}{M}F_{y}^{k}P_{x}^{k}+\frac{G^{2}}{2M}(F_{y}^{k})^{2}  \notag \\
&&+V_{trap}+\gamma B_{0}F_{z}^{k}+\gamma B_{0}GC^{k}+\gamma
B_{0}G^2D^{k}\}  \notag \\
&&+\sum_{k<l}V_{int}(\mathbf{r}_{k}-\mathbf{r}_{l})
\end{eqnarray}%
The atomic interaction takes the form \cite{Ho,Ohmi}
\begin{equation*}
V_{int}(\mathbf{r})=(c_{0}+c_{2}\mathbf{F}_{1}\cdot \mathbf{F%
}_{2})\delta (\mathbf{r})
\end{equation*}%
and $c_{0}=4\pi \hbar ^{2}(a_{0}+2a_{2})/3M,$ $c_{2}=4\pi
\hbar ^{2}(a_{2}-a_{0})/3M.$ The representation of the Hamiltonian in the
second quantized form reads
\begin{eqnarray}
\hat{H}_{eff} &=&\int d\mathbf{r\{}\hat{\Psi}_{i}^{\dag }[\frac{\mathbf{P}%
^{2}}{2M}\delta _{ij}-\frac{G}{M}P_{x}(F_{y})_{ij}+\frac{G^{2}%
}{2M}(F_{y}^{2})_{ij}  \notag \\
&&+V_{trap}+p_{0}(F_{z})_{ij}+p_0G C_{ij}+p_0G^2 D_{ij}]\hat{\Psi}_{j}
\notag \\
&&+\frac{c_{0}}{2}\hat{\Psi}_{i}^{\dag }\hat{\Psi}_{j}^{\dag }\hat{\Psi}_{j}%
\hat{\Psi}_{i}  \notag \\
&&+\frac{c_{2}}{2}\hat{\Psi}_{i}^{\dag }\hat{\Psi}_{i^{\prime }}^{\dag }%
\mathbf{F}_{ij}\cdot \mathbf{F}_{i^{\prime }j^{\prime }}\hat{\Psi}%
_{j^{\prime }}\hat{\Psi}_{j}\}  \label{hamone}
\end{eqnarray}%
where repeated indices are to be summed over and $\hat{\Psi}(\mathbf{r})(%
\hat{\Psi}^{\dag }(\mathbf{r}))$ is the field operator that annihilates
(creates) an atom in the \textit{i-}th hyperfine states with $i=1,0,-1$ at
location $\mathbf{r}$, and $p_{0}=\gamma B_{0}$ with $\gamma $ the
gyromagnetic ratio of the bosonic atoms.

The dynamics of the condensate components reveal a rich 
coupling between the spin and spatial degrees of freedom resulting in a
variety of interesting phenomena, including spin mixing, spin domain formation
and spin textures \cite{MSChang}. The internal and external dynamics 
are both very sensitive to the external magnetic
fields and field gradients and they can be decoupled  under certain conditions,
in particular, when the available spin dependent interaction is insufficient to
create spatial spin structures in the condensates. This occurs when the spin
healing length is larger than the size of the condensate, which allows us to 
focus on the coherent spin mixing oscillation of the spin populations. 
Practically we choose a proper field gradient $G$ to induce an energy 
in the same magnitude of spin interaction term $c_2$. The gradient will
flip the spin of atoms in the condensate but keep the three components still miscible 
and free of spin texture ($G<2\text{cm}%
^{-1}$ as in the MIT experiment \cite{Stenger}). As a result we can still
safely adopt the single spatial mode approximation (SMA) in the following \cite%
{Law,Pu,MSChang,Lee,Yi}.

We take
\begin{equation}
\hat{\Psi}_{i}^{\dag }=\sqrt{N}\phi (\mathbf{r})\hat{a}_{i}^{\dag }
\label{ansatz}
\end{equation}%
with $\phi (\mathbf{r})$ is defined by the Gross-Pitaevskii equation through
the spin-independent part $\hat{H}_{0}$%
\begin{equation}
\left( -\frac{\hbar ^{2}\mathbf{\nabla }^{2}}{2M}+V_{trap}+c_{0}N\left\vert
\phi \right\vert ^{2}\right) \phi =\hat{H}_{0}\phi =\mu \phi
\end{equation}%
where $N$ is the total number of the atoms and $\mu $ is the mean field
energy or the chemical potential. The $\hat{a}_{i}^{\dag }(\hat{a}_{i})$ is
the spin component operator that annihilates (creates) an atom with spin
\textit{i }$\left( i=1,0,-1\right) .$ Substitute the ansatz (\ref{ansatz})
into the Hamiltonian (\ref{hamone}) and neglect the spin-independent part,
our model then reads
\begin{equation}
H_{eff}=-\epsilon (\hat{a}_{1}^{\dagger }\hat{a}_{-1}\text{+}\hat{a}%
_{-1}^{\dagger }\hat{a}_{1})\text{+}\epsilon \hat{a}_{0}^{\dagger }\hat{a}%
_{0}\text{+}c_{2}\mathbf{\hat{F}}^{2}-p\hat{F}_{z}  \label{hameff}
\end{equation}%
with
\begin{eqnarray}
\mathbf{\hat{F}}^{2} &=&\hat{F}_{z}^{2}+(\hat{F}_{+}\hat{F}_{-}+\hat{F}_{-}%
\hat{F}_{+})/2  \notag \\
\hat{F}_{\pm } &=&\sqrt{2}(\hat{a}_{\pm 1}^{\dagger }\hat{a}_{0}+\hat{a}%
_{0}^{\dagger }\hat{a}_{\mp 1})  \notag \\
\hat{F}_{z} &=&\hat{a}_{1}^{\dagger }\hat{a}_{1}-\hat{a}_{-1}^{\dagger }\hat{%
a}_{-1}.
\end{eqnarray}%
The parameter $\epsilon =G^{2}/4M$ characterizes the spin-flipping process 
induced by the field gradient, and the spin interaction parameter is scaled as $c'_{2}=(c%
_{2}/2)\int d\mathbf{r}\left\vert \Phi (r)\right\vert ^{4}$ and we keep the original 
notation $c_2$ for ease of representation. The term $%
C_{ij}$ vanishes due to the fact that in ground state $\phi $ is a symmetric function
($\int d\mathbf{r}\phi ^{\ast }x\phi =0$ and $\int d\mathbf{r}\phi ^{\ast
}z\phi =0, etc.)$, so does the term $P_{x}(F_{y})_{ij}$ because
\begin{equation*}
\int d\mathbf{r}\phi ^{\ast }P_{x}\phi =\frac{M}{i\hbar }\int d\mathbf{r}%
\phi ^{\ast }[x,H_{0}]\phi =0.
\end{equation*}%
The term $D_{ij}$ amounts to a shift of the linear Zeeman energy $p=p_{0}+\tilde{p}$ with 
$\tilde{p}=p_0 G^{2} \int d\mathbf{r%
}\phi ^{\ast }x^{2}\phi $.

The $\epsilon $ term in Hamiltonian (\ref{hameff}) denotes the process that
flips the spin from $1$ to $-1$ or vise versa, and it plays the same role as the hopping term
of scalar condensates in a double well. We want to emphasize that this
term in the form of $\hat{a}_{i}^{\dag }(F_{y}^{2})_{ij}\hat{a}_{j}$ induces
the oscillation of the $z$-component magnetization $m$ and goes against the
quadratic Zeeman effect term $\hat{a}_{i}^{\dag }(F_{z}^{2})_{ij}\hat{a}_{j}$
in Refs. \cite{MSChang,Lee}, which instead adheres to the constancy of the
magnetization $m$ and keep the system in the polar phase
\cite{Stenger}. In a uniform field, the $z$-component magnetization $m$ is
a constant, and the system can be described by two canonical conjugate
variables: the population on spin-$0$ component $\rho _{0}$ and
the relative phase $\theta =\theta _{1}+\theta _{-1}-2\theta _{0}.$ The
system can be finally reduced to a nonrigid pendulum model \cite{MSChang}. Our model
system conserves only the total number of atom and another relative
phase $\theta ^{\prime }=\theta _{1}-\theta _{-1}$ arises. We will mainly
study the dynamics of the population on spin-$0$ component $\rho
_{0}(t)=N_{0}(t)/N$, and the  $z$-component magnetization $m(t)=(N_{1}(t)-N_{-1}(t))/N$. 
The populations on $\pm$ components are related to $\rho_0$ and $m$ through 
$\rho _{1}+\rho _{-1}+\rho _{0}=1$ and $m=\rho_{1}-\rho _{-1}$.

\section{The semiclassical model}

Instead of considering a Hilbert space of the Fock states $\left\vert
N_{1},N_{0},N_{-1}\right\rangle $, where the population dynamic can be
describe as
\begin{equation*}
\rho _{0}=\left\langle \psi _{i}\right\vert e^{i\hat{H}t/\hbar }\hat{a}%
_{0}^{\dagger }\hat{a}_{0}e^{-i\hat{H}t/\hbar }\left\vert \psi
_{i}\right\rangle /N
\end{equation*}%
with $\left\vert \psi _{i}\right\rangle $ an initial state which can be
taken as one of the basis in the Hilbert space, here we consider the
condensate to be in a coherent state, associated with a macroscopic wave
function with both magnitude and phase. In the study of spin mixing dynamics
\cite{Pu,Lee,WXZhang10}, it is customary to replace the operator $\hat{a}%
_{i}^{\dag }(\hat{a}_{i})$ by $c$ numbers $a_{i}^{\ast }(a_{i}),$ and the
coherent state is analogous to a classical field of complex amplitude with a
definite phase in each spin component
\begin{equation*}
\left\vert \Phi \right\rangle =\left\vert a_{1},a_{0},a_{-1}\right\rangle
=\left\vert \sqrt{N_{1}}e^{i\theta _{1}},\sqrt{N_{0}}e^{i\theta _{0}},\sqrt{%
N_{-1}}e^{i\theta _{-1}}\right\rangle
\end{equation*}%
Semiclassical equations of motion can be derived from Hamiltonian (\ref%
{hameff}) as%
\begin{eqnarray}
i\hbar \dot{a}_{1} &=&2c_{2}(\tilde{F}_{z}a_{1}+\tilde{F}_{-}a_{0}/\sqrt{2}%
)-\epsilon a_{-1}-pa_{1}  \notag \\
i\hbar \dot{a}_{0} &=&\sqrt{2}c_{2}(\tilde{F}_{+}a_{1}+\tilde{F}%
_{-}a_{-1})+\epsilon a_{0}  \label{dynamic} \\
i\hbar \dot{a}_{-1} &=&2c_{2}(-\tilde{F}_{z}a_{-1}+\tilde{F}_{+}a_{0}/\sqrt{2%
})-\epsilon a_{+1}+pa_{-1}  \notag
\end{eqnarray}%
where the quantities $\tilde{F}_{\pm },\tilde{F}_{z}$ are c number
counterparts of the operators $\hat{F}_{\pm }$ and $\hat{F}_{z}$
respectively. The neglected spin-independent part of the Hamiltonian (\ref%
{hamone}) give each of the three equations in Eq. (\ref{dynamic}) a constant
energy shift $\mu $ that can be trivially eliminated by changing $a_{i}$ to $%
a_{i}e^{-i\mu t/\hbar }$. For simplicity we rescale the
time as $t\rightarrow \left\vert c_{2}\right\vert t/\hbar$. The three coupled
Gross-Pitaevskii equations (\ref{dynamic}) for the interacting condensate
amplitudes $a_{i}^{\ast }(a_{i})$ describe the dynamics in terms of the
inter-component phase difference and population imbalance.

\begin{figure}[tbp]
\includegraphics[width=3.5in]{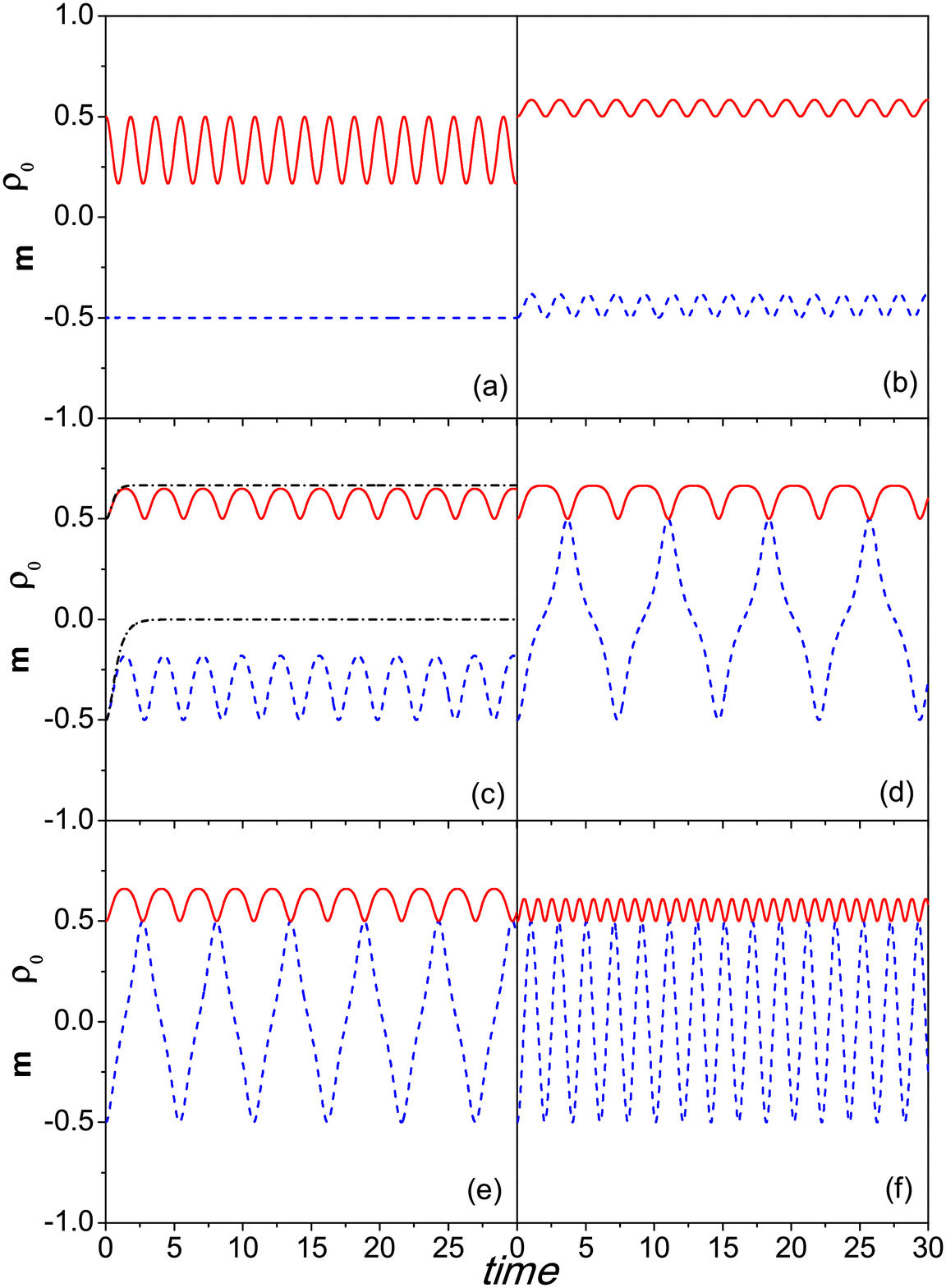}
\caption{(Color online) The dependence of the dynamics of $\protect\rho %
_{0}(t)$ (red solid) and $m(t)$ (blue dashed) on the parameters of $\protect%
\epsilon $ at fixed values of $p=0,$ c$_{2}=1$ (in units of $\left\vert
c_{2}\right\vert $), and $\protect\epsilon =0$ (a)$,1.45$ (b)$,1.49,1.50$ (c)%
$,$ $1.51$ (d), $1.55$ (e) and $2.25$ (f). Fig.1(c) shows the critical
transition parameters of $\protect\rho _{0}(t),m(t)$ with $\protect\epsilon $
=$1.50$ (black dashed dot line). Time is in units of $\left\vert c_{2}\right\vert t/\hbar$.}
\label{fig1}
\end{figure}
\begin{figure}[tbp]
\includegraphics[width=3.5in]{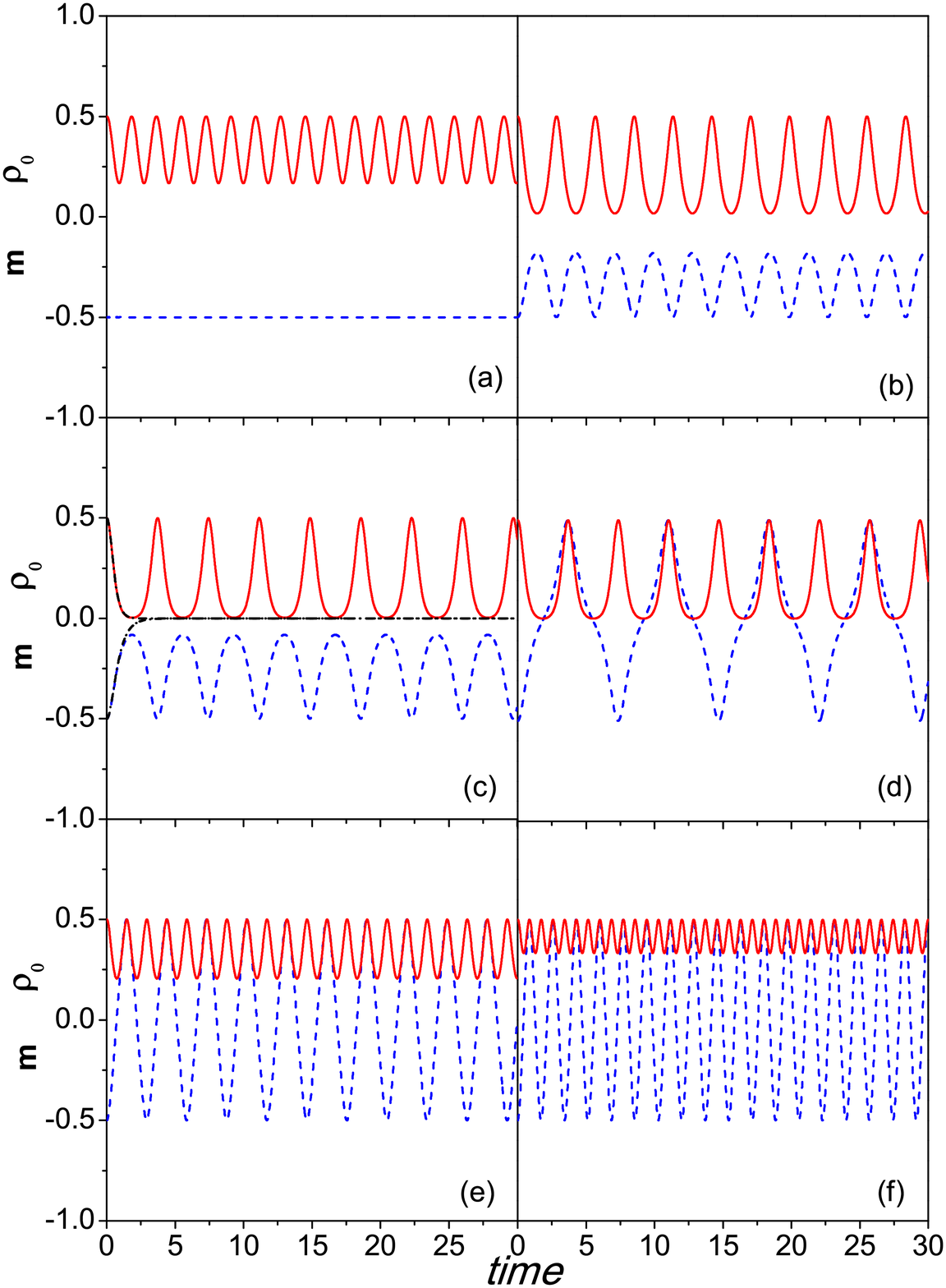}
\caption{(Color online) The dependence of the dynamics of $\protect\rho %
_{0}(t)$ (red solid) and $m(t)$ (blue dashed) on the parameters of $\protect%
\epsilon $ at fixed values of $p=0,$ c$_{2}=-1$ (in units of $\left\vert
c_{2}\right\vert $), and $\protect\epsilon =0$ (a)$,0.45$ (b)$,0.49,0.50$ (c)%
$,0.51$ (d)$,$ $0.85$ (e), and $1.50$ (f). Fig.2 (c) shows the critical
transition parameters of $m(t)$ with $\protect\epsilon $ =$0.50$ (black
dashed dot line). Time is in units of $\left\vert c_{2}\right\vert t/\hbar$.}
\label{fig2}
\end{figure}

\section{Results and discussion}

First, we consider an initial distribution with imbalance between $%
\left\vert +1\right\rangle$ and $\left\vert -1\right\rangle$ components,
i.e. $\left\vert \Phi (0)\right\rangle =\left\vert 0,\sqrt{N/2}e^{i\theta
_{0}},\sqrt{N/2}e^{i\theta _{-1}}\right\rangle $ and illustrate the dynamics
of population $\rho _{0}(t)$ and magnetization $m(t)$ for both $c_{2}>0$ and
$c_{2}<0.$
\begin{figure}[tbp]
\includegraphics[width=3.75in]{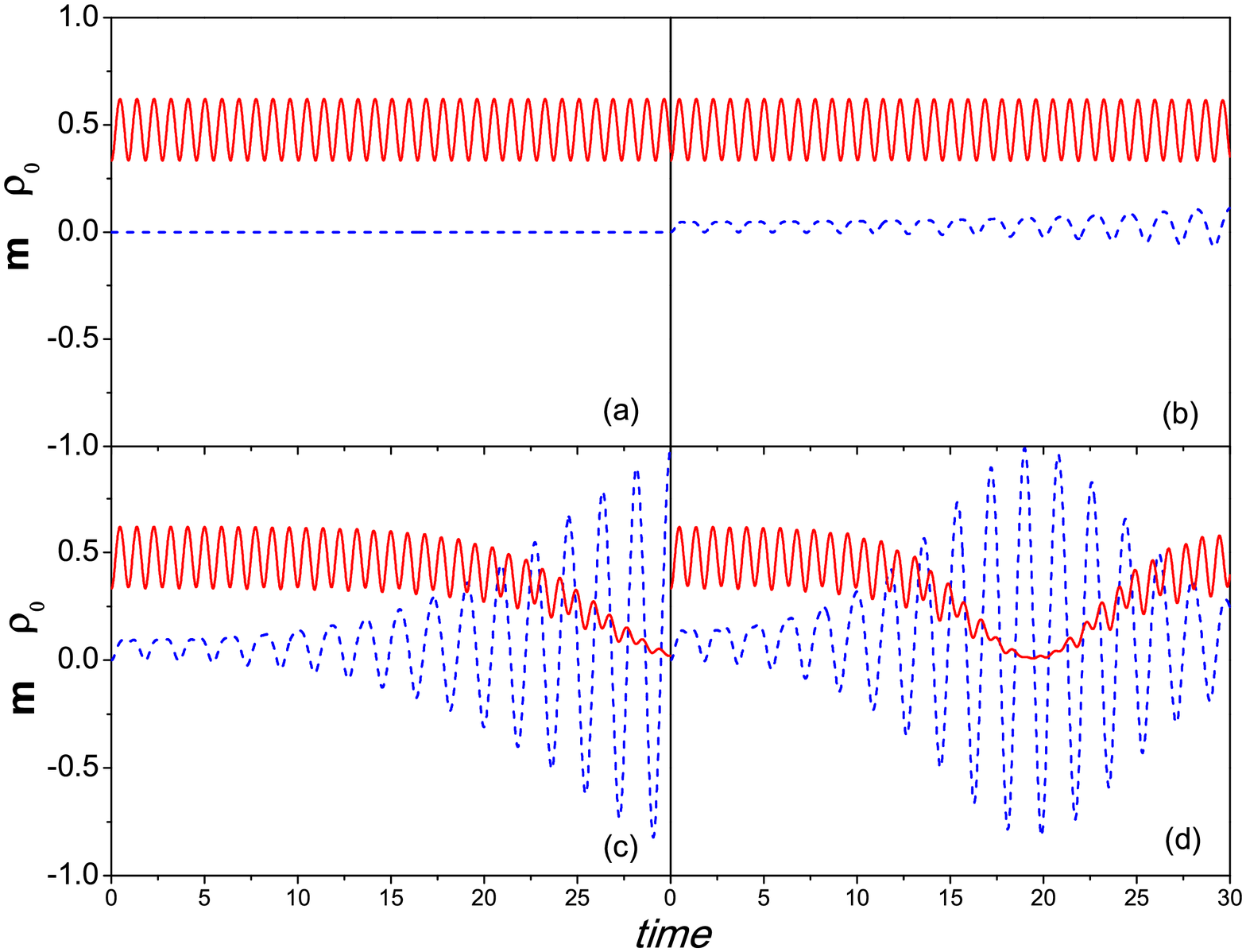}
\caption{(Color online) The dependence of the dynamics of $\protect\rho %
_{0}(t)$ (red solid) and $m(t)$ (blue dashed) on the parameters of $p$ at
fixed values of $\protect\epsilon =1.55,$ c$_{2}=1$ (in units of $\left\vert
c_{2}\right\vert $), and $p=0$ (a)$,0.2$ (b)$,0.4$ (c)$,$ $0.6$ (d).  Time is in units of $\left\vert c_{2}\right\vert t/\hbar$.}
\label{fig3}
\end{figure}
\begin{figure}[tbp]
\includegraphics[width=3.75in]{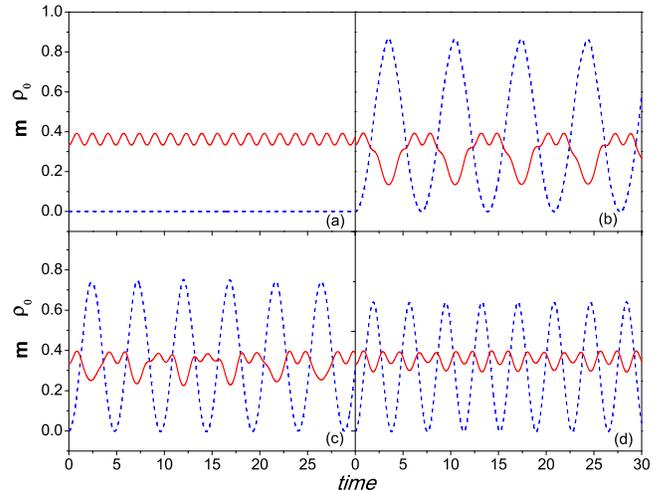}
\caption{(Color online) The dependence of the dynamics of $\protect\rho %
_{0}(t)$ (red solid) and $m(t)$ (blue dashed) on the parameters of $p$ at
fixed values of $\protect\epsilon =0.55,$ c$_{2}=-1$ (in units of $%
\left\vert c_{2}\right\vert $), and $p=0$ (a)$,0.2$ (b)$,0.4$ (c)$,$0.6 (d).
Time is in units of $\left\vert c_{2}\right\vert t/\hbar$.}
\label{fig4}
\end{figure}
In order to give prominence to the effect of the $\epsilon $ term, we
consider $p=0$ first, and set the phases $\theta _{0}=\theta _{-1}=0.$ This
initial imbalance $N_{1}(0)-N_{-1}(0)=-N/2$ provides a "Junction voltage"
and the magnetization oscillation was induced by $\epsilon$. In Fig. \ref%
{fig1}, we show the solutions of equations (\ref{dynamic}) for $c_{2}>0$ and
illustrative parameters $\epsilon =0,1.45,1.49,1.50,1.51,1.55$ and $2.25$,
in the unit of $\left\vert c_{2}\right\vert $, respectively. We find that at
the very beginning when $\epsilon $ is small the magnetization oscillates
with small amplitude around an equilibrium above the initial value of $%
m(0)=-0.5$ as in Fig.(\ref{fig1}b), which is analogous to the "macroscopic
quantum self-trapping" effect in double well system \cite{doublewell}.
Meanwhile the dynamics of $\rho _{0}(t)$ experiences a crossover from
sinusoidal to non-sinusoidal oscillation, with the population $\rho_0(t)$
averaged over time changing from less than the initial value $\rho
_{0}(0)=0.5$ to larger than it. As $\epsilon $ increases, there is a
critical transition for $\epsilon =1.50,$ black dashed line in Fig.(\ref%
{fig1}c), then the oscillation extends to the range between $-0.5$ and $0.5$%
. Accompanied by the arising of the "Josephson tunneling" \cite{doublewell}
of the magnetization, the dynamics of $\rho _{0}$ has been sufficiently
suppressed in Fig.(\ref{fig1}f). The coherent scattering of the internal
Zeeman components $2\left\vert 0\right\rangle \rightleftharpoons \left\vert
1\right\rangle +\left\vert -1\right\rangle $ was suppressed by the $\epsilon
$ term with the process $\left\vert 1\right\rangle \rightleftharpoons
\left\vert -1\right\rangle .$
\begin{figure}[tbp]
\includegraphics[width=3.5in]{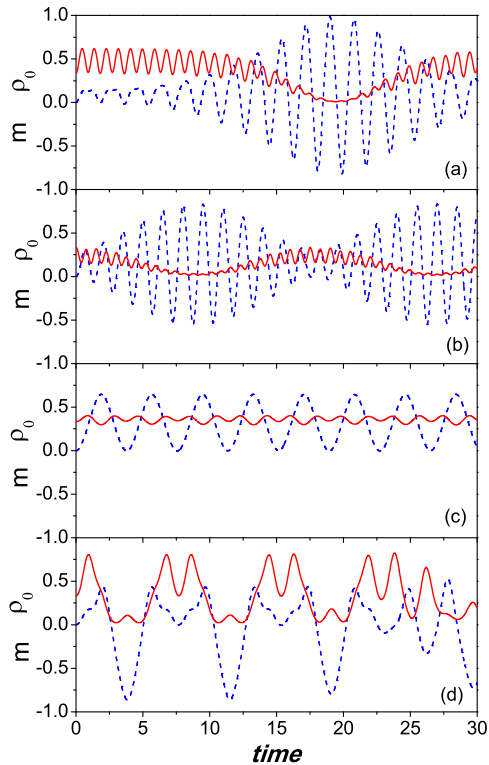}
\caption{(Color online) The dependence of the dynamics of $\protect\rho %
_{0}(t)$ (red solid) and $m(t)$ (blue dashed) on the phases of the initial
condition $\left\vert \Phi (0)\right\rangle =\left\vert \protect\sqrt{N/3}%
e^{i\protect\theta _{1}},\protect\sqrt{N/3}e^{i\protect\theta _{0}},\protect%
\sqrt{N/3}e^{i\protect\theta _{-1}}\right\rangle $ with parameters of $%
c_{2}=1,$ $\protect\epsilon =0.55,$ $p=0.6$ and $\protect\theta _{1}=\protect%
\theta _{0}=\protect\theta _{-1}=0$ for (a), $c_{2}=1,$ $\protect\epsilon %
=0.55,$ $p=0.6$ and $\protect\theta _{1}=\protect\theta _{-1}=\protect\pi %
/2, $ $\protect\theta _{0}=\protect\pi $ for (b), $c_{2}=-1,$ $\protect%
\epsilon =0.55,$ $p=0.6$ and $\protect\theta _{1}=\protect\theta _{-1}=%
\protect\theta _{0}=0$ for (c), $c_{2}=-1,$ $\protect\epsilon =0.55,$ $p=0.6$
and $\protect\theta _{1}=\protect\theta _{-1}=\protect\pi /2,$ $\protect%
\theta _{0}=\protect\pi $ for (d). Time is in units of $\left\vert c_{2}\right\vert t/\hbar$.}
\label{fig5}
\end{figure}
The critical behavior depends on $\epsilon $, as can be easily found from
the energy conservation and the extreme point for the minimization of the
energy. Considering an arbitrary wave function $\left\vert \Phi
\right\rangle =\left\vert xe^{i\theta _{1}},ye^{i\theta _{0}},ze^{i\theta
_{-1}}\right\rangle ,$ the relative average energy of the system when $p=0$
can be described as
\begin{eqnarray}
E &=&c_{2}[2y^{2}(x^{2}+z^{2}+2xz\cos \theta)+(x^{2}-z^{2})^{2}]
\notag \\
&&-2\epsilon xz\cos \theta '+\epsilon y^{2}  \label{EEE}
\end{eqnarray}%
with $\theta =\theta _{+}+\theta _{-1}-2\theta _{0},$ and $\theta'
=\theta _{+}-\theta _{-1}.$ For $c_{2}>0,$ the critical point favors
that $y=\sqrt{2/3},x=z=\sqrt{1/6},$ $\theta=\pi ,\theta'=\pi .$
According to the energy conservation condition $E_{c}=E_{initial},$ we can
get the critical value $\epsilon _{c}/c_{2}=1.50.$

For $c_{2}<0,$ we illustrate the dynamics for different parameters in Fig.(%
\ref{fig2}) with $\epsilon =0,0.45,0.49,0.50,0.51,0.85$ and $1.50,$ in the
unit of $\left\vert c_{2}\right\vert $ respectively. We find that the
oscillation of $m(t)$ here is almost the same as in the polar interaction
case $c_2>0$. On the other hand, $\rho _{0}(t)$ is not suppressed at the
beginning (Fig.(\ref{fig2}b)), instead, it was enhanced. When $\epsilon
>0.5, $ and the oscillation of $\rho _{0}$ is still active until $\epsilon $
reaches $1$ or even large value. This ferromagnetic feature is quite
different from the $c_{2}>0$ case. The latter case shows some "repulsive"
effect between the Zeeman components. The critical value $\epsilon
_{c}/(-c_{2})=0.5$ is derived analytically through the equation (\ref{EEE})
with the extreme point $y=0,x=z=\sqrt{1/2},$ $\theta=\theta'=0.$

Next, we consider a balanced initial distribution with $\left\vert \Phi
(0)\right\rangle =\left\vert \sqrt{N/3}e^{i\theta _{1}},\sqrt{N/3}e^{i\theta
_{0}},\sqrt{N/3}e^{i\theta _{-1}}\right\rangle $ and $\theta _{1}=\theta
_{-1}=\theta _{0}=0$, where the "Junction voltage" between the $\left\vert
1\right\rangle $ and $\left\vert -1\right\rangle $ components vanishes. We
consider the effect of the parameter $p,$ while we choose a fixed value of $%
\epsilon =1.55$ (in the unit of $\left\vert c_{2}\right\vert $) for $c_{2}>0$
and $\epsilon $ =0.55 for $c_{2}<0$ as shown in Fig.(\ref{fig3}) and Fig.(%
\ref{fig4}).

We find that the $p$ term acts as a switch for the dynamics of the
magnetization $m(t),$ and the cases for $c_{2}>0$ and $c_{2}<0$ are quite
different. For $c_{2}>0$ in Fig.(\ref{fig3}), the fast oscillation of $m(t)$
is modulated by a beat frequency, and, when the amplitude of the envelope
function reaches its maximum the population of $\rho _{0}$ is completely
suppressed to zero. For the $c_{2}<0$ case, the oscillation of $m(t)$ is
also induced by $p$, but the magnetization is always positive. The dynamics
of $\rho _{0}$ shows completely anharmonic behavior with the amplitude first
enhanced then reduced as $p$ increases, and no beat frequency modulation of
the $m(t)$ occurs.

However, specific features of quantum nature cannot be addressed
satisfactorily within a mean-field treatment. In an early experiment 
\cite{MSChang2} on an $F=1$ $^{87}$Rb condensate, atoms all 
prepared initially in the state $\left\vert 0,N,0\right\rangle $ are 
observed to exhibit a damped oscillation
accompanied by large fluctuations during the spin-mixing
evolution. The dynamics of a polar initial state $\left\vert 0,N,0\right\rangle $ 
is trivial, i.e. the population $\rho_0$ remains a constant, within a mean-field treatment, 
but many-body quantum dynamics shows interesting damped oscillation \cite{Law}.
The presence of a field gradient will enlarge the Hilbert space in quantum treatment
due to the failure of the conservation of magnetization $m$ and the related calculation 
on this feature will be published elsewhere.

Finally, let's consider the effect of the phase difference between the three
components. Fig.(\ref{fig5}) shows the dynamics in the presence of an
initial phase difference for both $c_{2}>0$ and $c_{2}<0$ cases. We find
that the influence of the phase difference for both cases are obvious. The
beat frequency oscillation of the magnetization remains in the case of $%
c_{2}>0$ with a shift of the envelop center, but for the $c_{2}<0$ the phase
difference changes the amplitude of $m(t)$ which extends down to the
negative part of the axis and the dynamics of $\rho _{0}$ is also altered
drastically.

\section{Conclusion}

The dynamics of spin-1 BEC in a nonuniform magnetic field is studied with
the emergence of an additional spin-flipping term
induced by the field gradient, which has an effect to reverse the spin from $+1$ to $%
-1$ and vise versa. Due to this spin flipping process the system magnetization $m(t)$ is
not a constant any more, instead, it shows characteristic oscillation
identical to that of a scalar BEC in double well. Meanwhile, the dynamics of $\rho
_{0}$ was greatly altered. We present the dynamics of $\rho _{0}(t)$ and $%
m(t)$ under different initial conditions and the effect of phase difference
is also shown. We find that the results for the polar ($c_{2}>0$) and
ferromagnetic ($c_{2}<0$) cases are quite different. The small magnetic
field gradient is chosen properly to give rise to the flipping of the spin
between $+1$ and $-1$ but still keep the three components miscible. These results
highlight the possibility to manipulate the coherent dynamics of the spinor condensate with a field
gradient, which is accessible to the current experimental techniques.

This work is supported by the NSF of China under Grant No.
11074153, the National Basic Research Program of China (973 Program) under
Grant No. 2011CB921601, the NSF of Shanxi Province, Shanxi
Scholarship Council of China, and the Program for New Century Excellent
Talents in University (NCET).


\begin{thebibliography}{99}
\bibitem{Stamper-Kurn} D. M. Stamper-Kurn, M. R. Andrews, A. P. Chikkatur,
S. Inouye, H.-J. Miesner, J. Stenger, and W. Ketterle, Phys. Rev. Lett.
\textbf{80}, 2027 (1998).

\bibitem{Law} C. K. Law, H. Pu, and N. P. Bigelow, Phys. Rev. Lett. \textbf{%
81}, 5257 (1998).

\bibitem{Pu} H. Pu, C. K. Law, S. Raghavan, J. H. Eberly, and N. P. Bigelow,
Phys. Rev. A \textbf{60}, 1463 (1999); H. Pu, S. Raghavan, and N. P.
Bigelow, \textit{ibid.} \textbf{61}, 023602 (2000).

\bibitem{MSChang} M.-S. Chang, Q. S. Qin, W. X. Zhang, L. You, and M. S.
Chapman, Nat. Phys. \textbf{1}, 111 (2005); W.X. Zhang, D. L. Zhou, M.-S.
Chang, M.S. Chapman, and L. You, Phys. Rev. A \textbf{72}, 013602 (2005).

\bibitem{MSChang2} M.-S. Chang, C. D. Hamley, M. D. Barrett, J. A. Sauer,
K.M. Fortier, W. Zhang, L. You, and M. S. Chapman, Phys. Rev. Lett. \textbf{%
92}, 140403 (2004).

\bibitem{Cheng} R. Cheng, J.-Q. Liang and Y. Zhang, J. Phys. B: At. Mol.
Opt. Phys. \textbf{38}, 2569 (2005).

\bibitem{Lee} L. Chang, Q. Zhai, R. Lu, and L. You, Phys. Rev. Lett. \textbf{%
99}, 080402 (2007); Q. Zhai, L. Chang, R. Lu, and L. You, Phys. Rev. A
\textbf{79}, 043608 (2009).

\bibitem{WXZhang10} W.X. Zhang, B. Sun, M. S. Chapman, and L. You, Phys.
Rev. A \textbf{81}, 033602 (2010).

\bibitem{Ho} T.-L. Ho, Phys. Rev. Lett. \textbf{81}, 742 (1998).

\bibitem{Ohmi} T. Ohmi and K. Machida, J. Phys. Soc. Jpn. \textbf{67}, 1822
(1998).

\bibitem{Stenger} J. Stenger, S. Inouye, D. M. Stamper-Kurn, H.-J. Miesner,
A. P. Chikkatur and W. Ketterle, Nature (London) \textbf{396}, 345 (1998).

\bibitem{domain} L. E. Sadler, J. M. Higbie, S. R. Leslie, M. Vengalattore,
and D. M. Stamper-Kurn, Nature (London) \textbf{443}, 312 (2006).

\bibitem{HoYip} T.-L. Ho and S.-K. Yip, Phys. Rev. Lett. \textbf{84}, 4031
(2000).

\bibitem{Mueller06} E. J. Mueller, T.-L. Ho, M. Ueda, and G. Baym, Phys.
Rev. A \textbf{74}, 033612 (2006).

\bibitem{Ciobanu} C. V. Ciobanu, S.-K. Yip, and T.-L. Ho, Phys. Rev. A
\textbf{61}, 033607 (2000).

\bibitem{Ueda} M. Koashi and M. Ueda, Phys. Rev. Lett. \textbf{84}, 1066
(2000); M. Ueda and M. Koashi, Phys. Rev. A \textbf{65}, 063602 (2002).

\bibitem{Ho3} R. B. Diener and T.-L. Ho, Phys. Rev. Lett. \textbf{96},
190405 (2006); L. Santos and T. Pfau, \textit{ibid} \textbf{96}, 190404
(2006); H. M\"{a}kel\"{a} and K.-A. Suominen, Phys. Rev. A \textbf{75},
033610 (2007).

\bibitem{Xuone} Z. F. Xu, Y. Zhang, and L. You, Phys. Rev. A \textbf{79},
023613 (2009).

\bibitem{Xutwo} Z. F. Xu, J. Zhang, Y. Zhang, and L. You, Phys. Rev. A
\textbf{81}, 033603 (2010).

\bibitem{zj} J. Zhang, Z. F. Xu, L. You and Y. Zhang, Phys. Rev. A \textbf{82%
}, 013625 (2010).





\bibitem{doublewell} A. Smerzi, S. Fantoni, S. Giovanazzi and S.-R. Shenoy,
Phys. Rev. Lett. \textbf{79}, 4950 (1997).

\bibitem{HoShenoy} T.-L. Ho and V. B. Shenoy, Phys. Rev. Lett. \textbf{77}, 2595 (1996).

\bibitem{Yi} S. Yi, \"{O}. E. M\"{u}stecaplioglu, C. P. Sun, and L. You,
Phys. Rev. A \textbf{66}, 011601(R) (2002).

\end{thebibliography}
\end{document}